\documentclass[11pt,sort&compress]{elsarticle}

\makeatletter
  \long\def\pprintMaketitle{\clearpage
  \iflongmktitle\if@twocolumn\let\columnwidth=\textwidth\fi\fi
  \resetTitleCounters
  \def\baselinestretch{1}%
  \printFirstPageNotes
  \begin{center}%
 \thispagestyle{pprintTitle}%
   \def\baselinestretch{1}%
    {\large\bf\@title}\par\vskip5pt
    \normalsize\elsauthors\par\vskip5pt
    \footnotesize\itshape\elsaddress\par\vskip10pt
    \end{center}%
  \gdef\thefootnote{\arabic{footnote}}%
  }
\makeatother

\newcommand\blfootnote[1]{%
  \begingroup
  \renewcommand\thefootnote{}\footnote{#1}%
  \addtocounter{footnote}{-1}%
  \endgroup
}

\journal{}
\usepackage{amsmath}
\usepackage{amsfonts}
\usepackage{graphicx}
\usepackage{braket}
\usepackage{qcircuit}
\usepackage{subfigure}
\usepackage{float}
\usepackage{xspace}
\usepackage[margin=1in]{geometry}
\usepackage{booktabs}
\usepackage{xspace}
\usepackage{epsf}
\usepackage{setspace}
\usepackage{abstract}

\usepackage[]{siunitx}
\usepackage[
    protrusion=true,
    activate={true,nocompatibility},
    final,
    tracking=true,
    kerning=true,
    spacing=true,
    factor=1100]{microtype}
\SetTracking{encoding={*}, shape=sc}{40}

\usepackage{xcolor}
\definecolor{lightblue}{rgb}{0.63, 0.74, 0.78}
\definecolor{seagreen}{rgb}{0.18, 0.42, 0.41}
\definecolor{orange}{rgb}{0.85, 0.55, 0.13}
\definecolor{silver}{rgb}{0.69, 0.67, 0.66}
\definecolor{rust}{rgb}{0.72, 0.26, 0.06}

\colorlet{lightsilver}{silver!30!white}
\colorlet{darkorange}{orange!75!black}
\colorlet{darksilver}{silver!65!black}
\colorlet{darklightblue}{lightblue!65!black}
\colorlet{darkrust}{rust!85!black}

\usepackage{bm}
\newcommand{\ve}[1]{\bm{#1}}

\newcommand{\bu}{\ve{u}}

\newcommand{\br}{\ve{r}}

\newcommand{\bc}{\ve{c}}

\newcommand{\eps}{\varepsilon}

\newcommand\Rey{\mbox{\text{Re}}\xspace}

\usepackage{hyperref}
\hypersetup{
  colorlinks=true,
}

\usepackage[labelfont=bf,font=small]{caption}

\usepackage[nameinlink]{cleveref}
\crefname{equation}{}{}

\usepackage[parfill]{parskip}

\begin{document}

\hypersetup{
  linkcolor=darkrust,
  citecolor=seagreen,
  urlcolor=darkrust,
  pdfauthor=author,
}

\begin{frontmatter}

\title{{\large\bfseries A multiple-circuit approach to quantum resource reduction \\ with application to the quantum lattice Boltzmann method}}

\author[1]{Melody Lee}
\author[2]{Zhixin~Song}
\author[1]{Sriharsha~Kocherla}
\author[3]{Austin~Adams}
\author[4]{Alexander~Alexeev}
\author[1,4,5]{Spencer~H.~Bryngelson}
\ead{shb@gatech.edu}

\address[3]{School of Computational Science \& Engineering, Georgia Institute of Technology, Atlanta, GA 30332, USA\vspace{-0.15cm}}
\address[2]{School of Physics, Georgia Institute of Technology, Atlanta, GA 30332, USA\vspace{-0.15cm}}
\address[3]{School of Computer Science, Georgia Institute of Technology, Atlanta, GA 30332, USA\vspace{-0.15cm}}
\address[4]{George W.\ Woodruff School of Mechanical Engineering, Georgia Institute of Technology, Atlanta, GA 30332, USA\vspace{-0.15cm}}
\address[5]{Daniel Guggenheim School of Aerospace Engineering, Georgia Institute of Technology, Atlanta, GA 30332, USA}

\date{}

\end{frontmatter}

\begin{abstract}
    This work proposes a multi-circuit quantum lattice Boltzmann method (QLBM) algorithm that leverages parallel quantum computing to reduce quantum resource requirements.
    Computational fluid dynamics (CFD) simulations often entail a large computational burden on classical computers.
    At present, these simulations can require up to trillions of grid points and millions of time steps.
    To reduce costs, novel architectures like quantum computers may be intrinsically more efficient for these computations.
    Current quantum algorithms for solving CFD problems are based on single quantum circuits and, in many cases, use lattice-based methods.
    Current quantum devices are adorned with sufficient noise to make large and deep circuits untenable.
    We introduce a multiple-circuit algorithm for a quantum lattice Boltzmann method (QLBM) solve of the incompressible Navier--Stokes equations.
    The method, called QLBM-frugal, aims to create more practical quantum circuits and strategies for differential equation-based problems.
    The presented method is validated and demonstrated for 2D lid-driven cavity flow.
    The two-circuit algorithm shows a marked reduction in CNOT gates, which consume the majority of the runtime on quantum devices.
    Compared to the baseline QLBM technique, a two-circuit strategy shows increasingly large improvements in gate counts as the qubit size, or problem size, increases.
    For $64$ lattice sites, the CNOT count was reduced by 35\%, and the gate depth decreased by 16\%.
    This strategy also enables concurrent circuit execution, further halving the seen gate depth.
\end{abstract}

\blfootnote{
\noindent Code available at: \url{https://github.com/comp-physics/QLBM-frugal}
}

\section{Introduction}\label{s:introduction}

\subsection{Motivation and application of quantum computation for scientific computing}

Computationally intensive algorithms are obstacles to the scientific computing and engineering communities.
Quantum computers can achieve complexity improvements by leveraging quantum mechanics properties, making them of interest.
A prominent example of this so-called \textit{quantum advantage} is demonstrated by Shor's algorithm~\citep{Shor_1997}.
The algorithm finds the prime factorization of integers in logarithmic time, an exponential speedup over the optimal classical algorithm.
Other quantum algorithms include linear systems solvers~\citep{harrowQuantumAlgorithmSolving2009, clader2013preconditioned}, Monte Carlo methods~\citep{montanaro2015quantum}, and machine learning algorithms~\citep{havlivcek2019supervised, liu2021rigorous}.
With the recent progress of quantum hardware, these algorithms have been applied in industries involving optimization~\citep{farhi2014quantum}, quantum chemistry~\citep{chem1}, and finance~\citep{herman2023quantum}.
The potential for such large speedups makes partial differential equation (PDE) solvers attractive.
However, classical problems in this realm tend to be nonlinear and non-unitary, presenting hurdles for quantum computing~\citep{itani2}.
Attention has been turned toward addressing this challenge.

\subsection{Quantum computation for computational fluid dynamics}

Computational fluid dynamics (CFD) problems often require some of the world's largest classical supercomputers to solve.
Focus has been directed toward alternative quantum algorithms for these problems~\citep{gourianov2022quantum,succi2023quantum,succi2024ensemble,steijl2019quantum,moawad2022investigating,steijl2018parallel,gaitan2020finding,bharadwaj2020quantum,bharadwaj2023hybrid,li2023potential,oz2022solving,jaksch2023variational,givi2020quantum,xu2018turbulent,xu2019quantum,sammak2015quantum}.
The connection between quantum mechanics and fluid dynamics emerged early on through the Madelung equations, a reformulation of the Schr\"{o}dinger equations~\citep{Madelung_1926,succi2023quantum}
This relationship between quantum computing and CFD was further explored in 1993, when \citet{Succi_Benzi_1993} used the lattice Boltzmann equation to describe non-relativistic quantum mechanics.

Over a decade later, the Harrow--Hassidim--Lloyd (HHL) algorithm, a sparse linear solver with quantum advantage, was used to solve the incompressible Navier--Stokes equations~\citep{harrowQuantumAlgorithmSolving2009}.
Many quantum CFD solvers are also based on the quantum singular value transformation algorithm~\citep{Gilyen_Su_Low_Wiebe_2019}.
Hybrid quantum and classical algorithms have also been presented for CFD problems~\citep{lapworth2022hybrid,Bharadwaj_Sreenivasan_2023,Song_Deaton_Gard_Bryngelson_2024}.
One reduces the CFD problem to a quantum-solvable form, and the quantum device solves the linear system.
However, the advantage posed by the HHL algorithm does not account for the state preparation and readout requirements~\citep{Aaronson_2015}.
Implementing efficient input and output methods for quantum computers is in itself a challenge: loading an exponentially large state space representative of $n$-qubits requires $2^n$ amplitudes be somehow encoded on a quantum computer.
Various approaches have been taken, but present generalized encoding methods are upper-bounded by an exponential complexity~\citep{Niemann_Datta_Wille_2016,Mozafari_Soeken_Riener_DeMicheli_2020,Araujo_Park_Petruccione_daSilva_2021}, with slight improved efficiency and depth on encoding sparse matrices~\citep{Gleinig_Hoefler_2021}.

\subsection{Quantum computing for mesoscale CFD methods}

This paper extends bodies of work focused on mesoscale methods for fluid simulations.
The dissipative particle dynamics and Boltzmann equation-based methods are commonly used in classical computation.
The Boltzmann equation (which characterizes the statistical behavior of a system of particles) is focused on herein.
This equation is primarily solved using lattice gas automata~\citep{lga2,lga1} and the lattice Boltzmann method (LBM)~\citep{lbmbook, clbm1}.
LBM is the more advantageous of the two, because it enables noise resilience and flexibility in complex multiphysics simulations.
Prior bodies of work have presented adaptations of both methods.
For example, previous bodies of work proposed quantum algorithms for the lattice gas model~\citep{yepezLatticeGasQuantumComputation1998, yepezQuantumLatticegasModel2001, yepezQuantumLatticegasModel2002}.
More recently, a quantum lattice gas model was presented by \citet*{Zamora_Budinski_Niemimaki_Lahtinen_2024} with improved efficiency.
In 2020, \citet{todorova2020quantum} described a quantum algorithm for the collisionless Boltzmann equation.
This method accounts for the case of nearly free molecular flows and ignores complex particle collisions.

Linearizing the problem is the first step in mapping the lattice Boltzmann method to a quantum-suitable form.
This struggle with nonlinear terms is due to the linear framework on which quantum computing is defined.
The Carleman and Koopman--von~Neumann embeddings have been used to linearize equations for quantum computing~\citep{Ito_Tanaka_Fujii_2023}.
The Carleman method embeds the variables of nonlinear equations directly into a quantum state.
This approach is advantageous over the Koopman--von~Neumann method, which instead embeds the variables through orthogonal polynomials.
\citet{itani} demonstrated a Carleman linearization for the collision term and later extended the work to demonstrate unitary evolution for both the collision and streaming operators~\citep{itani2}.
These approaches were extended in later work~\citep{sanavio2023quantum,todorova2020quantum}.

\citet{budinskiquantumalgorithmadvection2021} employed the Carleman linearization technique to introduce a quantum lattice Boltzmann method (QLBM) for the advection--diffusion equation.
The proposed method simulates particles across a grid of lattice sites and extracts macroscopic quantities from mesoscopic simulations.
Since the resultant matrix is not necessarily Hermitian, the system must be discretized in terms of time~\citep{harrowQuantumAlgorithmSolving2009,Childs_Kothari_Somma_2017}.
The linearized system of equations is then solved via a quantum linear system algorithm~\citep{Ito_Tanaka_Fujii_2023}.
\citet{sanavio2023quantum} introduced a QLBM algorithm that simulated Kolmogorov-like flows using the Carleman linearization method.
They showed that their implementation of the single time-step operator would have a fixed circuit depth.
\citet{budinski2021quantum} adapted the quantum circuit approximation using the stream function--vorticity formulation of the Navier--Stokes equations.
This adaptation removes the pressure term, reducing the problem to the advection--diffusion equation and Poisson equation, solved on a single circuit.

\subsection{Limitations of quantum computing}

Many of the proposed algorithms are presumed to operate on fault-tolerant quantum computers~\citep{budinski2021quantum, sanavio2023quantum, gaitan2020finding, Costa_Jordan_Ostrander_2019}.
Although the theory is promising, noise and decoherence on present quantum devices result in high error rates.
Consequently, foundational algorithms, such as HHL, are unable to be supported on modern quantum hardware~\citep{lapworth2022hybrid}.
The coherence time of a device is the range of time during which the quantum device reliably maintains its state such that the results obtained are within reason~\citep{Lidar_Chuang_Whaley_1998, Ladd_2010}.
Circuits whose runtime exceeds this coherence time output solutions that contain too much noise to be useful.
The sensitivity to the environment makes it difficult to scale quantum algorithms, as they require large numbers of qubits.
As a result, the aforementioned quantum CFD algorithms must be simulated.

The limitations in current quantum hardware is denoted Noisy Intermediate-Scale Quantum (NISQ). 
These limitations have driven research towards hybrid variational quantum algorithms (VQAs) for both adiabatic and gate-based quantum computing~\citep{Jaksch_Givi_Daley_Rung_2023}.
The adiabatic configuration models the system as an energy minimization problem cast as an Ising Hamiltonian~\citep{Srivastava_Sundararaghavan_2019}.
However, this approach is unable to adapt to allow control over precision and other parameters~\citep{Ray_Banerjee_Nadiga_Karra_2019}.

Gate-based VQAs are proposed as near-term replacements for HHL without the quantum advantage.
\citet{demirdjian2022variational} solved the 1D advection--diffusion equation via Carleman linearization with reasonable accuracy.
This approach uses a variational quantum linear solver~\citep{bravo-prieto_variational_2020} to compute the solution.
\citet{lubasch2020variational} demonstrated VQAs may account for nonlinear effects.
However, beyond a certain circuit depth in noisy devices, VQAs are unlikely to outperform optimal classical counterparts on combinatorial problems~\citep{DePalma_Marvian_Rouzé_França_2023}.
Thus, methods to reduce the circuit depth and number of quantum gates applied are beneficial.

\subsection{Distributed and parallelized quantum systems}

One approach to reducing quantum resource use is distributed computation.
A distributed quantum computer is a novel architecture composed of a network of devices.
Its interactions occur via quantum teleportation, physical transport, or a predefined classical method.
Attention has been dedicated toward nonlocal quantum operations that enable their communication~\citep{Yimsiriwattana_Lomonaco_2005, Buscemi_2012, Sarvaghad_2021}.
Distributed computing is of interest in quantum communications, quantum internet, and quantum key distribution research, among others~\citep{Gisin_Thew_2007,Villegas_2024,Lee_2014}.
This architecture has been conceptualized as a technique to adapt to limitations on small capacity quantum computers, including on the foundational Simon's and Grover's algorithms~\citep{Yimsiriwattana_Lomonaco_2005, Cuomo_2020,Tan_2022}.
Sub-processes within the algorithms are delegated among the devices.

There has also been interest in parallelized quantum processes.
In this architecture, multiple circuits are run concurrently on a single device.
\citet{Broadbent_Kashefi_2009} devises a technique for parallelizing quantum circuits that reduces polynomial-depth circuits to logarithmic depth.
The quantum processor's qubits are allocated to minimize interference from the concurrent processes.
\citet{Matteo_Mosca_2016} demonstrated that this method reduces gate depth and speedup in select cases.
As such, investigating the parallelization of proposed quantum algorithms is beneficial when solving larger problems.

\subsection{Objectives}

This work formulates a two-circuit 2D~QLBM algorithm, QLBM-frugal, to solve the incompressible Navier-Stokes equations.
The method is validated against the classical LBM solution to the 2D~lid-driven cavity flow problem.
The quantum resources required for QLBM-frugal are compared to other approaches, and the parallelized implementation is discussed.
\Cref{s:classical} describes the classical LBM theory for simulating the advection--diffusion equation.
\Cref{s:mqlbm} shows a single-circuit quantum implementation of the LBM for solving the advection--diffusion equation.
In \cref{s:qlbm}, the two-circuit quantum implementation for solving the 2D~Navier--Stokes equations is provided.
\Cref{s:results} presents verification against the classically-solved LBM simulation and quantum resource estimations.
\Cref{s:limitations} discusses the current limitations of the work.
\Cref{s:conclusions} summarizes the contributions of the presented method and its potential impact.

\section{Mathematical formulation}\label{s:classical}

\subsection{Advection--diffusion equation}

The advection--diffusion equation describes flow in a way that treats advection and diffusion as concurrent processes.
The governing advection--diffusion PDE is
\begin{gather}
   \frac{\partial \phi}{\partial t} + c\frac{\partial \phi}{\partial x} = D\frac{\partial^2 \phi}{\partial x^2},
   \label{e:ad}
\end{gather}
where advection is defined by $c({\partial \phi}/{\partial x})$ and diffusion is defined by $D({\partial^2 \phi}/{\partial x^2})$.
Here, $c$ and $D$ are the advection and diffusion coefficients, and $\phi(t,x)$ is a scalar concentration field.
We use the advection--diffusion algorithm devised by \citet{budinskiquantumalgorithmadvection2021} as the foundation for the presented multi-circuit approach.

\subsection{General lattice Boltzmann formulation}

The LBM model for fluid flow simulates the evolution of particle flow over time, given by~\citep{Rothman1997LatticeGasCA,rivet_boon_2001} as
\begin{gather}
    f_\alpha(\br+e_\alpha \Delta t, t+\Delta t) = (1-\epsilon)f_\alpha(\br,t) + \epsilon f_\alpha^{\text{(eq)}} + \Delta w_\alpha S,
    \label{e:boltzmann}
\end{gather}
where $f_\alpha$ indicates the particle distribution along each $\alpha$.
$e_\alpha$ is the velocity of a particle in link $\alpha$, $t$ is time, $\delta t$ is the time step, $\br$ is the cell position, $S$ is the source term, $w_\alpha$ is the proportion of particles streaming in link $\alpha$, and $\epsilon = \Delta t/\tau$ where $\tau$ is the relaxation time.
We more generally define $\br$ using Cartesian coordinates, whose notation depends on the dimension of the vector space.
For example, in a one-dimensional space, we express $\br = (x)$, while in a two-dimensional space, we express $\br = (x, y)$. 
\Cref{f:lbm} shows how the particle distribution may be visualized along a grid of cells.
Each time step indicates some particle movement into neighboring cells in accordance with the above equation.

To solve \cref{e:boltzmann} in the one- and two-dimensional planes, we investigate three configurations of the spatial lattice model.
For a D$N_\mathrm{site.}$Q$N_\mathrm{links}$ configuration, we consider a $N_\mathrm{site.}$-dimensional lattice grid with $N_\mathrm{links}$ links, where each link is a direction in which particles are propagated.
This paper considers advection--diffusion results for the D$1$Q$2$, D$1$Q$3$, and D$2$Q$5$ configurations.

\subsection{Formulation for one-dimensional scheme}

Initially, we devise a one-dimensional model for the D$1$Q$2$ configuration.
Two velocity vectors are considered such that
\begin{gather}
    e_1 = -e_2 = \Delta x / \Delta t
\end{gather}
for corresponding distribution functions $f_1$ and $f_2$.
The symmetric boundary condition of the LBM imposes this equality.
Our model resolves around the local equilibrium distribution function~\citep{mohamad2011lattice}, formulated as
\begin{gather}
    f_\alpha^{\text{(eq)}}(\br,t) =
        w_\alpha \phi(\br,t)
        \left(1 + \frac{\boldsymbol{e_\alpha} \cdot \bc}{c_s^2}\right),
    \label{e:feq}
\end{gather}
where $\phi(\br, t)$ corresponds to the lattice site at lattice position $\br$ = $(x_i)$, $\bc$ is the advection velocity vector, and $c_s$ is the speed of sound.
We set the weights $w_{1, 2} = 0.5$ and $c_s = 1$, in accordance with the approach by \citet{budinskiquantumalgorithmadvection2021}.
Likewise, the D$1$Q$3$ configuration is considered with three velocity vectors, where $e_0 = 0$ and $e_1 = -e_2 = 1$. Let the speed of sound $c_s = 1/\sqrt{3}$ following \citet{servan2009non}.

\subsection{Formulation for two-dimensional scheme}

We now consider the two-dimensional D$2$Q$5$ scheme.
\citet{budinski2021quantum} shows that these equations are reduced to resemble the advection--diffusion formulation, referencing his earlier work~\citep{budinskiquantumalgorithmadvection2021} in setting the parameters and governing formulas.
We adapt this advection--diffusion formulation to a multiple-circuit method for advection and diffusion.
This formulation is expressed as
\begin{gather}
   \frac{\partial \phi}{\partial t} + c\frac{\partial \phi}{\partial x} = D\frac{\partial^2 \phi}{\partial x^2},
   \label{e:ad}
\end{gather}
where advection is defined by $c({\partial \phi}/{\partial x})$ and diffusion is defined by $D({\partial^2 \phi}/{\partial x^2})$.
Here, $c$ and $D$ are the advection and diffusion coefficients and $\phi(t,x)$ is a scalar concentration field.

\begin{figure}[h!]
    \centering
    \includegraphics{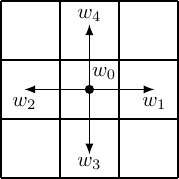}
    \caption{
        Illustration of a D2Q5 LBM lattice with a streaming particle and link weights $w_\alpha$.
    }
    \label{f:lbm}
\end{figure}

The equilibrium distribution function is solved using \cref{e:feq}, where $\br = (x_i, y_i)$. 
Let $c_s = 1/\sqrt{3}$~\citep{servan2009non} and the weights be defined such that $w_0 = 2/6$ and $w_{2,3,4} = 1/6$ respective to the directions illustrated in \cref{f:lbm}~\citep{ginzburg2005equilibrium}.

Following standard practice~\citep{lbmbook},
the relaxation time $\tau$ relates to the diffusion constant $D$ with
\begin{gather}
    D = c_s^2 \left(\tau - \frac{\Delta t}{2}\right).
\end{gather}
Herein we use $D = 1/6$ and $\tau = \Delta t$. Recall we have set $c_s = 1/\sqrt{3}$.
This gives $\eps = \tau / \Delta t  = 1$ and
\begin{gather}
    f_\alpha(\br+\boldsymbol{e_\alpha} \Delta t, t+\Delta t) = f_\alpha^{\text{(eq)}}
    \label{e:adrex}
\end{gather}
simplifies as
\begin{equation}
    f_\alpha(\br+\boldsymbol{e_\alpha} \Delta t, t+\Delta t) = w_\alpha \phi(\br,t)\left(1 + \frac{\boldsymbol{e_\alpha} \cdot \bc}{c_s^2}\right),
    \label{e:adlinks}
\end{equation}
which includes the collision \cref{e:feq} and the streaming steps \cref{e:adrex}.
The concentration field is thus
\begin{equation}
    \phi(\br,t) = \sum_{\alpha = 0}^{N-1} f_\alpha(\br,t),
    \label{e:admacro}
\end{equation}
summing the particle distributions across all link directions $\alpha$.

\section{Quantum Lattice Boltzmann method}\label{s:mqlbm}

\subsection{Quantum Lattice Boltzmann method for the advection--diffusion equation}\label{s:advectiondiffusion}

We now discuss the premise of the body of work devised by \citet{budinski2021quantum}.
Consider, first, the advection--diffusion equations.
The quantum circuit performs the collision operator, followed by the streaming of particles and recalculation of macroscopic variables.
Boundary conditions are applied at the end of each time step.
No special care is needed when boundary conditions are periodic; the left and right shift gates $L$ and $R$ automatically propagate boundary conditions to the designated lattice site for each link $\alpha$.
\Cref{fig:advCircuit} illustrates how the quantum circuit's qubits are organized into 4~registers, or groupings of qubits, to solve the advection--diffusion equation using a D2Q5 scheme.
Registers $r_0$ and $r_1$ contain $\log N_\mathrm{site.}$ qubits, where $N_\mathrm{site.}$ is the number of lattice sites in each dimension.
Register $d$ has $\lceil \log_2 Q \rceil$ qubits, where $Q$ is the number of LBM links $\alpha$.
Register $a$ holds a single ancilla qubit necessary for applying a non-unitary collision operator.

\subsubsection{Encoding input}\label{ss:encoding}

For the current study, the amplitude encoding technique described in the work of \citet{shende06} is employed.
Amplitude encoding is part of the Qiskit toolkit~\citep{Qiskit}, albeit also a generic encoding algorithm with an expensive gate count.
We further elaborate on this in \cref{s:limitations}.

At the start of each time step, we are given a distribution $\phi(\br, t)$
We define $\phi(\alpha, \br)$ to be the value of $\phi$ at time $t$ and position $\br$ at link direction $\alpha$.
Given a discretized concentration
\begin{gather}
    \phi(\alpha, \br) = [\phi(0,0), \phi(0,1),\phi(0,2), \dots, \phi(4,M-2),\phi(4, M-1),\phi(4,M)],
    \label{e:init}
\end{gather}
the initial statevector $\ket{\psi_0}$ is
\begin{gather}
    \ket{\psi_0} = \ket{0}_a \otimes \frac{1}{\vert\vert \phi \vert\vert} \sum_{i=0}^{2M-1} \phi(\alpha, \br)\ket{i}.
    \label{e:initstate}
\end{gather}
This formulation normalizes $\phi(\alpha, \br)$, and the initial qubit states follow from amplitude encoding.
In \cref{fig:advCircuit}, register $r_0$ stores the data at each link direction $\alpha$.
This data can be retrieved by specifying the link $\alpha$ in register $d$.

Applying the collision operator to the initial statevector $\ket{\psi_0}$ is equivalent to multiplying the $\phi(\alpha,\br)$ with weight coefficients
\begin{gather}
    w_\alpha \left(1 + \frac{\boldsymbol{e_\alpha} \cdot \bc}{c_s^2}\right),
    \label{e:coeff}
\end{gather}
which follow from \cref{e:feq}.
This strategy is discussed further in the next subsection.

\subsubsection{Collision operator}\label{ss:colision}

\begin{figure}
    \centering
    \includegraphics{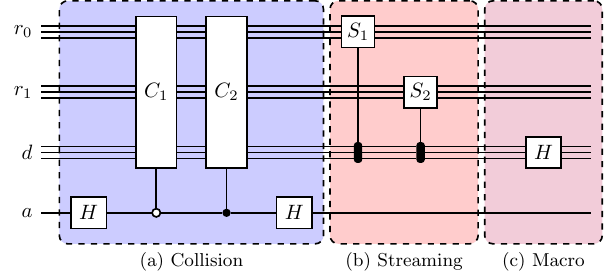}
    \caption{
        QLBM circuit for an advection--diffusion time step as devised by \citet{budinskiquantumalgorithmadvection2021}.
        The collision operator $A = (C_1 + C_2)/2$, and shift operators $R$, $L$, $D$, and $U$ propagate particles in each link direction $\alpha$.
        In panel~(c), a Hadamard~\citep{mikeike} gate is applied to the qubits in register $d$.
    }
    \label{fig:advCircuit}
\end{figure}

The collision step (\cref{fig:advCircuit}~(a)) computes the equilibrium distribution function $f_a^{\text{(eq)}}$, which requires computing the proportion of the distribution $\phi$ in each link $\alpha$.
The collision operator entails applying the coefficient matrix $A$ to the current statevector $|\psi_0\rangle$.

The coefficient matrix $A$ is not unitary, so it cannot be directly translated into a quantum gate.
To mitigate this, an important strategy, devised by \citet{Childs2012HamiltonianSU}, is employed by \citet{budinski2021quantum}. Suppose we split $A$ into a linear combination of unitary matrices, $C_1$ and $C_2$, related to the original matrix as
\begin{gather}
    C_{1,2} = A \pm i \sqrt{I-A^2}.
    \label{e:c2}
\end{gather}
As $A = (C_1 + C_2)/2$, an operation with $A$ is computed via block encoding~\citep{blockencoding1, blockencoding2}, where $C_1$ and $C_2$ are unitary, but $A$ is, in general, not.

\begin{figure}[h]
    \centering
    \includegraphics{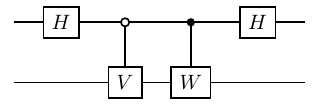}
    \caption{
        A block encoding of $B$ = $(V+W)/2$.
        Here, $V$ and $W$ are unitary matrices.
    }
    \label{fig:block-encoding}
\end{figure}

The circuit in \cref{fig:block-encoding} evolves an input statevector $\ket{\psi}$ according to the unitary
\begin{gather}
    U = \frac{1}{2}\begin{bmatrix}V + W & V - W \\ V - W & V + W \end{bmatrix},
    \label{e:blockenc}
\end{gather}
where $B = (V+W)/2$.
Thus, $B$ is a subblock of the block matrix $U$.

Post-selection selects quantum states for specific measurement outcomes.
Here, we use post-selection to measure the result after the collision operator for a $0$ ancilla (after the block encoding).

The collision matrix is
\begin{gather}
    A =
    \begin{bmatrix}
        k_1I_n  & 0 \\
        0       & k_2I_n
    \end{bmatrix},
    \quad \text{where}
    \quad
    k_{\alpha} = w_\alpha \left(1 + \frac{\boldsymbol{e_\alpha} \cdot \bc}{c_s^2}\right)
\end{gather}
are the link coefficients described by \cref{e:coeff}.
Panel~(a) of \cref{fig:advCircuit} uses a linear combination of unitaries to apply matrix $A$ to the input.
The Hadamard gates are used for the block encoding process along with $C_1$ and $C_2$ operations, which are derived from the unitary matrices of \cref{e:c2}.
The coefficients representing the proportion of particles in each link $\alpha$ are $k_1$ and $k_2$.
The collision matrices are
\begin{gather}
    C_{1,2} =
    \begin{bmatrix}
        \exp(\pm i\arccos{(k_1)})I_n & 0 \\
        0                            & \exp(\pm i\arccos{(k_2)})I_n
    \end{bmatrix}.
\end{gather}

The collision operator $A$ transforms statevector $\ket{\psi_0}$ via a linear combination of $C_1$ and $C_2$, but requires an ancilla qubit $a$, which stores orthogonal data $(C_1 - C_2)/2$ where the ancilla is $\ket{1}$.
The orthogonal data was ignored through post-selection.
The result of this linear combination is
\begin{gather}
    \ket{\psi_1} = \frac{1}{\vert\vert \phi \vert\vert} \sum_i a_{i,i}\phi_{i,i}\ket{i},
    \label{e:psi1}
\end{gather}
which encodes the post-collision values for each link direction $\alpha$, for which the ancilla is $\ket{0}$.

\subsection{Particle streaming}\label{ss:streaming}

\begin{figure}
    \centering
    \includegraphics{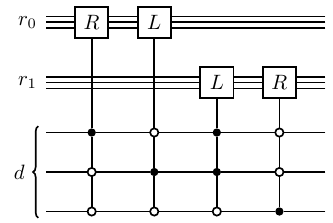}
    \caption{
        Streaming for a two-dimensional lattice grid.
        The streaming step uses right and left shift operators $R$ and $L$ to shift distributions in their respective link directions $\alpha_1$ and $\alpha_2$ by controlling the gates on qubits in each link register $d$.
    }
    \label{fig:streaming}
\end{figure}

The streaming step propagates particles in each link $\alpha$ to the neighboring site.
\Cref{fig:streaming} shows the shift operators $R_n$ and $L_n$, which are controlled on link qubits $d$ and stream particles to neighboring lattice sites.
The shift matrices are permutation matrices as
\begin{gather}
    R_n = \begin{bmatrix}
        0 & 0 & \cdots & 0 & 1 \\
        1 & 0 & \cdots & 0 & 0 \\
        0 & 1 & \cdots & 0 & 0 \\
        \vdots & \vdots & \ddots & \vdots & \vdots \\
        0 & 0 & \cdots & 1 & 0
    \end{bmatrix} \quad \text{and} \quad
    L_n = \begin{bmatrix}
        0 & 1 & \cdots & 0 & 0 \\
        \vdots & \vdots & \ddots & \vdots & \vdots \\
        0 & 0 & \cdots & 1 & 0 \\
        0 & 0 & \cdots & 0 & 1 \\
        1 & 0 & \cdots & 0 & 0
    \end{bmatrix},
    \label{e:shifts}
\end{gather}
and are both unitary matrices.

\begin{figure}
    \centering
    \subfigure[\small Left shift operator $L$.]{\includegraphics{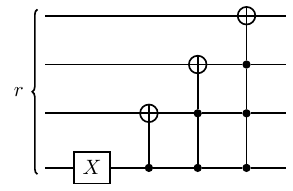}}
    \subfigure[\small Right shift operator $R$.]{\includegraphics{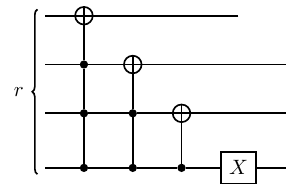}}
    \caption{
        Circuit decomposition for (a) left shift and (b) right shift operators.
        These gates are applied to qubits in $r_1$ for left and right shifts and qubits in $r_2$ for up- and down-shifts.
    }
    \label{fig:prop_circuits}
\end{figure}

The resulting statevector $\ket{\psi_2}$ has a distribution shifted to a neighboring lattice site determined by its link value $\alpha_i$.
\Cref{fig:prop_circuits} shows how the right and left shift gates $R$ and $L$ are controlled on the link qubits $d$ to act on the part of the statevector for distribution $\alpha$.
\Cref{fig:streaming} shows the up and down operators $U$ and $D$, which are implemented via the $L$ and $R$ operators applied via $r_1$.

\subsection{Macroscopic variable retrieval}\label{ss:macro}

We retrieve the distribution $\phi(\br,t)$ by summing $f_\alpha^\mathrm{(eq)}$ over the link directions $\alpha$.
\Cref{fig:advCircuit} shows how this is accomplished by applying Hadamard gates to each of the qubits in link registers $d$, as shown in \cref{fig:advCircuit}~(c)~\citep{mikeike}.
A Hadamard gate $H$ applied to a statevector $\ket{\psi}$, resulting in
\begin{gather}
    H|\psi\rangle = \frac{1}{\sqrt{2}}\begin{bmatrix}
        1 & 1 \\
        1 & -1
    \end{bmatrix}
    \begin{bmatrix}
        a \\
        b
    \end{bmatrix}
    = \frac{1}{\sqrt{2}}\begin{bmatrix}
        a + b \\
        a - b
    \end{bmatrix}
    = \frac{a + b}{\sqrt{2}}|0\rangle + \frac{a - b}{\sqrt{2}}|1\rangle.
\end{gather}
When the Hadamard gate is applied to a link qubit in $d$, it stores the sum of the two amplitudes as the $\ket{0}$ amplitude and the difference as the $\ket{1}$ amplitude.
Thus, the sum can be post-selected by ignoring $\ket{1}$ measurements.
With this post-selection, the Hadamard gates sum the distributions but introduce a factor of $1/\sqrt{2}$ per gate.
This pre-factor is post-processed out of the computation by multiplication of a factor of $\sqrt{2}^{\log N_\mathrm{links}}$ where $N_\mathrm{links} = |\alpha|$ is the number of link distributions.
At each time step, we retrieve the circuit result via state tomography, which must be used to extract the first $N_\mathrm{site.}$ lattice site elements.

\section{Quantum Lattice Boltzmann method for the Navier--Stokes equations}\label{s:qlbm}

\subsection{Stream function--vorticity formulation}\label{ss:streamfunction}

The incompressible 2D Navier--Stokes equations in Cartesian coordinates are
\begin{align}
    \frac{\partial u}{\partial t} + u\frac{\partial u}{\partial x} +
        v\frac{\partial u}{\partial y} &=
        -\frac{\partial p}{\partial x} +
        \frac{1}{\Rey} \left(\frac{\partial^2 u}{\partial x^2} +
        \frac{\partial^2 u}{\partial y^2}\right), \\
    \frac{\partial v}{\partial t} + u\frac{\partial v}{\partial x} +
        v \frac{\partial v}{\partial y} &=
        -\frac{\partial p}{\partial y} +
        \frac{1}{\Rey} \left(\frac{\partial^2 v}{\partial x^2} +
        \frac{\partial^2 v}{\partial y^2}\right),
\end{align}
where $u$ and $v$ are the velocity components in the $x$ and $y$ coordinate directions, $p$ is the pressure, and $\Rey$ is the Reynolds number, which is the ratio of inertial to viscous effects~\citep{batchelor1967introduction}.

Taking the curl of the above Navier--Stokes equations recasts them in the so-called vorticity--stream function formulation, removing the pressure term $p$ and yielding
\begin{align}
    \frac{\partial^2 \psi}{\partial x^2} + \frac{\partial^2 \psi}{\partial y^2} &= -\omega,  \label{ss:nsPoi} \\
    \frac{\partial \omega}{\partial t} + u\frac{\partial \omega}{\partial x} + v\frac{\partial \omega}{\partial y} &= \frac{1}{\Rey} \left(\frac{\partial^2 \omega}{\partial x^2} + \frac{\partial^2 \omega}{\partial y^2} \right).\label{ss:nsDiff}
\end{align}

In this formulation, \cref{ss:nsPoi} and \cref{ss:nsDiff} use vorticity $\omega$ and stream function $\psi$ instead of directional speeds $u$ and $v$.
The velocity vector is thus $\bu = \{u, v\}$.
The stream function relates to the directional velocities as
\begin{gather}
    \frac{\partial \psi}{\partial x} = u
    \quad \text{and} \quad
    \frac{\partial \psi}{\partial y} = -v.
\end{gather}
Thus, \cref{ss:nsPoi} is a Poisson equation in stream function $\phi$ and \cref{ss:nsDiff} is an advection--diffusion equation in vorticity $\omega$.

\subsubsection{Lattice-based representation}

With the stream function--vorticity formulation, the collision, streaming, and macro lattice stages follow as
\begin{align}
    f_\alpha ^ {\text{(eq)}}(\br,t) &=
        w_\alpha \omega(\br,t)
        \left(1 + \frac{\boldsymbol{e}_\alpha \cdot \bu}{c_s^2}\right), \\
    f_\alpha(\br+\boldsymbol{e}_\alpha \Delta t, t+\Delta t) &=
        f_\alpha^{\text{(eq)}}, \\
    \omega(\br,t) &= \sum_\alpha f_\alpha(\br,t).
\end{align}
These stages, buttressed via the circuits of \cref{s:advectiondiffusion}, enable the vorticity $\omega$ computation.

The equilibrium distribution function for the Poisson equation ($\nabla^2\psi = -\omega$) is $g_\alpha^{\text{(eq)}}(\br,t) = w_\alpha \psi(\br,t)$.
The streaming and macro steps match those of \cref{ss:streaming,ss:macro}, but the source term $S = -\omega$ is added during the collision step~(a).
Thus, the relaxation operator is
\begin{gather}
    g_\alpha(\br+\boldsymbol{e}_\alpha \Delta t, t+\Delta t) =
        g_\alpha^{\text{(eq)}} + \Delta w_\alpha S,
\end{gather}
and macro retrieval equation
\begin{gather}
    \psi(\br,t) = \sum_\alpha g_\alpha(\br,t).
\end{gather}

\subsubsection{Boundary conditions}\label{ss:bcs}

The validation problem for the proposed method is a 2D~lid-driven cavity flow.
The spatial domain is $\Omega \in x,y$ with lengths $L_x$ and $L_y$ and boundary $\partial \Omega$.
The stream function $\psi$ is constant along the boundaries, with $\psi = 0$ used here.
For the lattice Boltzmann method,
\begin{gather}
    \psi = \sum_{\alpha=0}^{N_\mathrm{links}-1} g_\alpha^{\text{(eq)}} = g_0 + g_1 + g_2 + g_3 + g_4, \quad \text{so,} \quad
    g_{\partial\Omega} = -\sum_{\alpha | \alpha \neq \partial\Omega} g_\alpha.
\end{gather}

Defining the vorticity expression \cref{ss:nsPoi} in terms of the stream function and expanding it in its Taylor series gives
\begin{gather}
    \omega_{i,N_\mathrm{qubit}} = - 2\left( \frac{\psi}{\Delta y^2} + \frac{U}{\Delta y}\right),
\end{gather}
along the boundaries $\partial\Omega$, where $U$ is wall-parallel velocity of the top wall.
For a stationary wall, $U = 0$.
The wall equilibrium distribution in the direction of the wall is
\begin{gather}
    g(x,y)_{\partial\Omega} = -{\sum_{\alpha | \alpha \neq \partial\Omega} g_\alpha} - 2 \left(\frac{-\psi}{\Delta y^2} + \frac{U}{\Delta y}\right).
    \label{eq:boundaries}
\end{gather}

To implement \cref{eq:boundaries}, the matrix
\begin{gather}
    B = \begin{bmatrix}
        0       & \cdots    & 0 \\
        \vdots  & I_{N_\mathrm{qubit}-2}   & \vdots \\
        0       & \cdots    & 0
    \end{bmatrix},
\end{gather}

is applied to the statevector $|\phi \rangle$, where $B$ is of size $N_\mathrm{qubit} \times N_\mathrm{qubit}$, $N_\mathrm{qubit} = \log(N_\mathrm{site.})$ in each dimension, and $I_{n}$ denotes the size $n$ identity matrix.
Setting $g_{\partial \Omega} = 0$, while retaining other distribution values, enforces the boundary condition for the stream function circuit $|\psi_{\text{b.c.}}\rangle = |0\rangle^{\otimes N_\mathrm{qubit}}$.

The vorticity circuit boundary conditions are applied via a linear combination of $B$ to account for its nonlinear nature.
This linear unitary combination is
\begin{gather}
    D_{1,2} = B \pm i\sqrt{I_{N_\mathrm{qubit}} - B^2},
\end{gather}
following \cref{ss:colision}.

\begin{figure}
    \centering
    \includegraphics{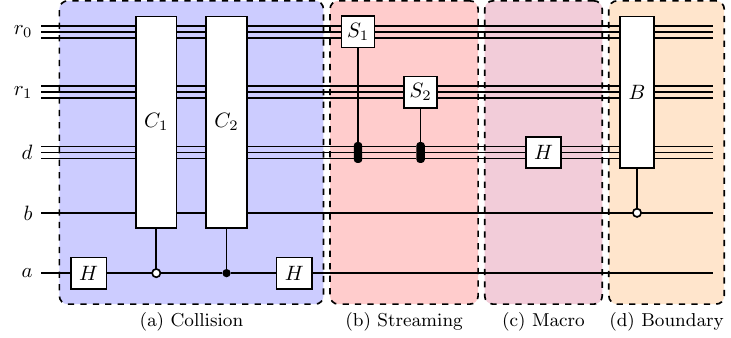}
    \caption{
        The original D2Q5 vorticity circuit proposed by \citet{budinskiquantumalgorithmadvection2021}, modified from the generalized circuit in \cref{fig:advCircuit} for boundary conditions.
        The changed circuit includes an additional qubit $b$ to store the boundary conditions and computations.
        We propose an algorithm that splits this circuit into two distinct parts.
    }
    \label{fig:vortCircuit}
\end{figure}

\subsection{Two-circuit model}

The deviation from \citet{budinski2021quantum}'s original algorithm lies in the two-circuit approach, which defines distinct circuits for the stream function and vorticity computation.
To devise this, we adopt the advection--diffusion circuits, as discussed in~\citep{budinskiquantumalgorithmadvection2021}, and apply different bounds specific to the lid-driven cavity case.
We discuss each circuit in more detail below.

\subsubsection{Vorticity circuit}\label{ss:vorticity}

\begin{figure}
    \centering
    \includegraphics{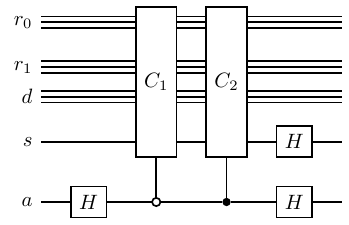}
    \caption{
        Collision portion of the stream function circuit.
        An additional qubit $s$ stores the source term, and an additional Hadamard gate on qubit $s$ after the block encoding adds the source term to the link distributions $\alpha_i$ before streaming.
    }
    \label{fig:streamCircuit}
\end{figure}

The vorticity circuit computes the vorticity $\omega$ for the current time step.
The vorticity circuit in \cref{fig:vortCircuit} and advection--diffusion circuit in \cref{fig:advCircuit} are nearly identical, save for the boundary conditions.
This matching occurs because the vorticity equation in \cref{ss:nsDiff} follows the same form as the diffusion equation.
The Navier--Stokes algorithm's vorticity circuit is the same as the advection--diffusion one if the boundary conditions are computed classically.

If boundary conditions are included, an extra qubit $b$ stores them.
Boundary conditions require additional computation, as enforcing such conditions is not a unitary operation.
For this, the linear combination of unitaries is used ~\citep{meyer1997quantum}, described further in \cref{ss:colision}.
The input to this circuit for the no-boundary version is the previous time step's vorticity $\omega_{t-1}$.
When using boundary conditions, the circuit input includes pre-computed boundary conditions in the boundary qubit $b$, computed following the descriptions in \cref{ss:bcs}.

\subsubsection{Stream function circuit}\label{ss:streamcircuit}

The stream function circuit is the other quantum circuit used in this two-circuit model.
The D2Q5 stream function circuit shown in \cref{fig:streamCircuit} is referred to for additional context.
This circuit is similar to the advection--diffusion circuit with classical boundary conditions but includes an additional source term.
\Cref{fig:streamCircuit} shows the key difference between these circuits, where the required additional qubit $s$ stores the source term, $S=-\omega$.

The $C_1$ and $C_2$ gates also operate on the source term, and the Hadamard gate on qubit $s$ adds this source to the stream function's equilibrium distribution function $g_\alpha^\mathrm{(eq)}$.
The same qubit-addition process from \Cref{ss:vorticity} applies.
Boundary conditions require an additional $b$ qubit store and a boundary gate $B$, which is not unitary and again is implemented via a linear combination of matrices following \cref{e:blockenc}.
The previous stream function, $\psi_{t-1}$, and source term, $S=-\omega_{t-1}$, serve as the input to the stream function circuit.
The boundary conditions are computed according to \cref{ss:bcs} for our quantum boundary condition variant.

\section{Simulations and results}\label{s:results}

The QLBM algorithm of \citet{budinski2021quantum} uses a single circuit to solve the Navier--Stokes equations.
This work follows from corresponding work on advection--diffusion algorithms~\citep{budinskiquantumalgorithmadvection2021}.
QLBM-frugal separates the computational process into distinct stream function and vorticity circuits.
We verify the algorithm against that of \citet{budinskiquantumalgorithmadvection2021}.
We consider two cases for the advection--diffusion algorithm, the D$1$Q$2$ and D$1$Q$3$ lattice schemes, and validate the implementation accuracy before separating of the stream function and vorticity circuits.
We also consider a lid-driven cavity problem to verify the two-circuit approach.
With this, we can compare how the resources use of the algorithm scale under increasing lattice site count against the work of \citet{budinskiquantumalgorithmadvection2021}.

\subsection{Advection--diffusion equation}\label{ss:advdiffusion}

We apply the QLBM circuit to the advection--diffusion equation to obtain results for two exemplar cases.
We verify these results against the expected outcomes returned by the classical LBM.
In the 1D case, D1Q2 and D1Q3 lattice schemes simulate a dense concentration of $\rho = 0.2$ at source $x_i = 10$, undergoing advection and diffusion with uniform advection velocity of $c = 1/5$ and diffusion coefficient $D = 1/6$.
The validation problem for the 2D case follows a diffusing concentration $\rho = 0.3$ at source $(x_i, y_j) = (4,4)$ and $0.1$ elsewhere, solved via a D2Q5 scheme.

\begin{figure}[htbp]
    \centering
    \begin{tabular}{c c }
        \includegraphics{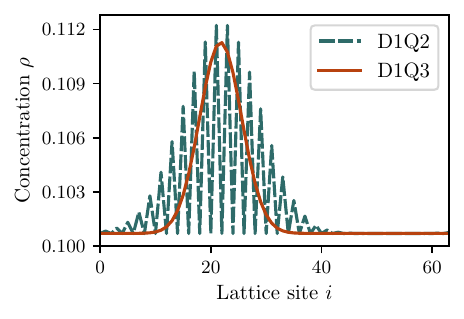} &
        \includegraphics{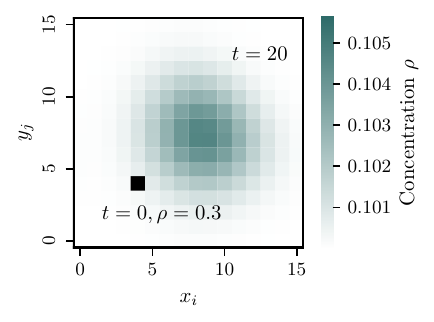} \\
        \small (a) D1Q2 and D1Q3 & \small (b) D2Q5
    \end{tabular}
    \caption{
        Quantum LBM (a) D1Q2, D1Q3, and (b) D2Q5 results for the advection--diffusion equation.
        The initial condition in (a) is a point source of $\rho = 0.2$ at $x=10$ and $\rho = 0.1$ otherwise.
        The initial condition in (b) is a point source of $\rho = 0.3$ at $(x,y) = (4,4)$ and $\rho = 0.1$ elsewhere.
    }
    \label{fig:advres}
\end{figure}

\Cref{fig:advres} shows the results of statevector simulations over $50$ timesteps. 
A checkboard pattern arises in the D1Q2 case because the distribution moves wholly into neighboring areas, resulting in half of the lattice sites having zero particles.
The algorithmic deficiency is remedied via the D1Q3 lattice scheme by setting the weight of vector $w_0$, specified in \cref{f:lbm}, to $2/3$.
\Cref{fig:advres}~(a) shows that D1Q3 resolves the checkerboarding problem of D1Q2.

\Cref{fig:advres}~(b) shows the results of the 2D test problem solved with the D2Q5 scheme after $20$ timesteps.
There exists an initial source at $(x_i = 4, y_j = 4)$, with an advection velocity in the positive $x$ and $y$ directions. 
The source advects in the direction of the velocity and diffuses outwards.
The solution behaves as expected and agrees with classical LBM results.

\begin{figure}
    \centering
    \includegraphics{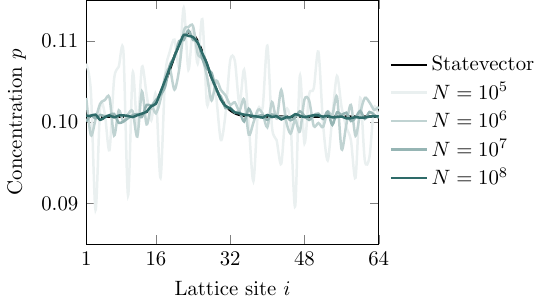}
    \caption{
        The results of the D$1$Q$3$ algorithm at iteration $50$ are obtained via finite sampling of the quantum circuit at various sample sizes $N_{\mathrm{samp.}}$.
    }
    \label{f:samplingresults}
\end{figure}

The results are further validated via finite sampling.
\Cref{f:samplingresults} displays the results of finite sampling with various shot sample sizes $N_{\mathrm{samp.}}$.
We verify the behavior of the algorithm by computing the state fidelity of results obtained via finite sampling to the statevector of the ideal solution.
The quantum state fidelity, $F_{\psi}(\rho)$, of a (probabilistic) mixed quantum state $\rho$ with respect to a pure quantum state $\psi$ is expressed as
\begin{equation}\label{e:fidelity}
    F_{\psi}(\rho) = \langle \psi | \rho | \psi \rangle = \textrm{Tr}(p|\psi \rangle \langle \psi |).
\end{equation}

\begin{figure}
    \centering
    \includegraphics{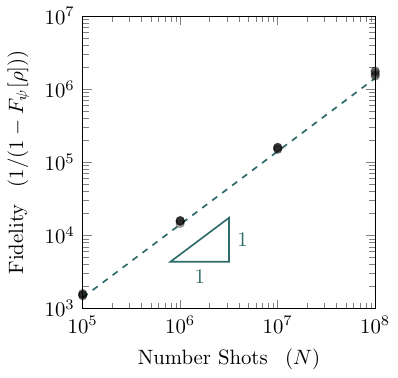}
    \caption{
        Verification via linear convergence with shot number of the advection--diffusion D1Q3 algorithm.
    }
    \label{f:fidelity}
\end{figure}

Here, $\psi$ is the expected solution obtained via the QLBM algorithm, and $\rho$ is the probabilistic outcome obtained via finite sampling methods.
From the equation defined in \cref{e:fidelity}, we conclude that if $\rho$ perfectly resembles the ideal solution, then we have $F_{\psi}(\rho) = 1$.
The fidelity is expected to be expressed as a function of the number of shots $N_{\mathrm{samp.}}$ used in the experiment, where in accordance to~\citep{Yu_Shang_Guhne_2022}
\begin{equation}
    N_\mathrm{samp.} \propto \frac{1}{1 - F_{\psi}(\rho)}
\end{equation}
\Cref{f:fidelity} shows the results of this verification.
The slope of the fit line in \cref{f:fidelity} is $1.01$, a 1\% difference from the expected proportion.

\subsection{Navier--Stokes equations}\label{ss:ns-results}

\begin{figure}
    \centering
    \includegraphics{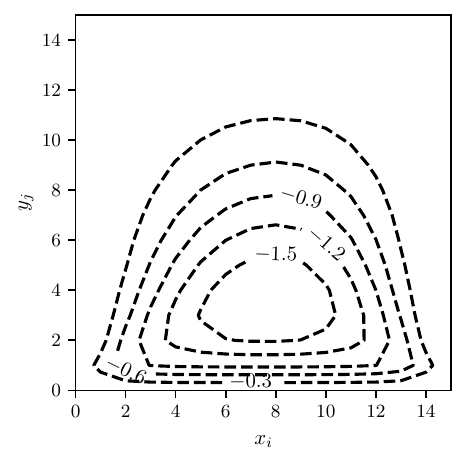}
    \caption{
        A 2D lid-driven cavity flow.
        Steady-state stream function isocontours are shown as labeled.
        The initial conditions are $\psi = 0$, $w = 0$, and $U = 1$, done over $80$ timesteps.
    }
    \label{fig:streamres}
\end{figure}

\Cref{fig:streamres} shows isocontours of the stream function for a lid-driven cavity problem.
The problem serves to verify the two-circuit QLBM against a classical implementation of the lattice Boltzmann method.

To define relative errors between the quantum and classical lattice algorithms, we denote
\begin{gather}
    \psi_{i,j} = \psi_{(x_i,y_j)} \quad \text{and} \quad \omega_{i,j} = \omega_{(x_i,y_j)},
\end{gather}
where $x_i = i \Delta x$ and $y_j = j \Delta y$.
The local $L_1$ relative error between the classical and two-circuit QLBM Navier--Stokes solver is, thus,
\begin{gather}
    \eps_{\psi;i,j} = \frac{\psi_{i, j}^{\text{classic.}} - \psi_{i, j}^{\text{quant.}}}{\psi_{i, j}^{\text{classic.}}} \quad \text{and} \quad
    \eps_{\omega;i,j} = \frac{\omega_{i, j}^{\text{classic.}} - \omega_{i, j}^{\text{quant.}}}{\omega_{i, j}^{\text{classic.}}}.
\end{gather}

\begin{tabular}{c c }
        \includegraphics{compound.pdf} &
        \includegraphics{d2q5res.pdf} \\
        \small (a) D1Q2 and D1Q3 & \small (b) D2Q5
    \end{tabular}

\begin{figure}
    \centering
    \begin{tabular}{c c }
        \includegraphics{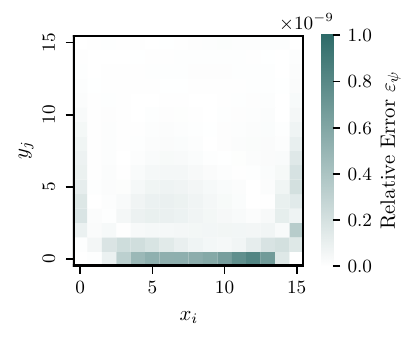} &
        \includegraphics{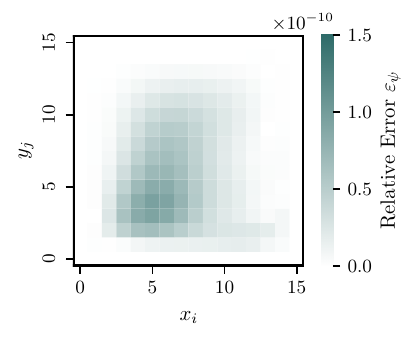} \\
        \small (a) Relative error $\eps_{\omega}$ for vorticity circuit & \small (b) Relative error $\eps_{\psi}$ for stream function circuit
    \end{tabular}
    \caption{
        Relative error between the classical LBM and two-circuit quantum methods for (a) vorticity and (b) stream function.
    }
    \label{fig:d2q5_circuit_comparison}
\end{figure}

The relative errors are shown for the cavity problem in \cref{fig:d2q5_circuit_comparison}.
\Cref{fig:d2q5_circuit_comparison} shows that the two-circuit QLBM agrees with the classical LBM when simulated with the statevector simulator in Qiskit's~SDK~\citep{Qiskit}.
The statevector simulator perfects exact tomography but is generally limited to small simulations due to exponentially increasing memory requirements with qubit numbers.

\subsubsection{Quantum resource estimation and improvement}

\begin{table}
    \centering
    \small
    \begin{tabular}{r c c}
     & CNOT Gates & Circuit Depth \\ \midrule
    Single-circuit QLBM & 25\phantom{.0} & 58\phantom{.0} \\
    Stream function & 4.3 & 9.4 \\
    Vorticity & 12\phantom{.0} & 39\phantom{.0} \\
    Stream function without boundaries & 4.2 & 15\phantom{.0} \\
    \end{tabular}
    \caption{Quantum resource estimation (all counts are in units of $10^4$) for a D2Q5 algorithm with lattice size $64 \times 64$.}
    \label{table:t}
\end{table}

Implementing the two-circuit method for solving the Navier--Stokes equations using quantum lattice-based algorithms shows quantum resource advantages over the single-circuit method.
The advantages are achieved in two areas: two-qubit gate count and runtime.
Two-qubit gates like CNOTs are slower and more error-prone than single-qubit gates, which has prompted bodies of work to reduce counts of these gates in quantum algorithms~\citep{Park_Ahn_2023}.
The projected runtime is calculated by summing the implementation time for each gate in the circuit.
As such, the runtime is approximately proportional to the circuit depth.

We show the reduction in gate counts by converting the circuit into a series of equivalent one- and two-qubit gates via the Qiskit transpiler.
This procedure is conducted on a $64 \times 64$ lattice, with no prior optimizations or conversions to backend-specific gate sets.
\Cref{table:t} show that pre-computing the boundary conditions using classical methods reduces the gate count by at least $35\%$. 
By running the QLBM-frugal stream function and vorticity circuits concurrently, circuit depth is reduced to the maximum depth of its constituent circuits.
\Cref{table:t} shows that a 33\% reduction in circuit depth when compared against the single-circuit algorithm by~\citet{budinski2021quantum}.

We similarly transpile the results relative to a select backend.
During compilation, we set the optimization level to be 3, as defined in \citet{Qiskit} documentation.
The optimization levels in Qiskit range from 0 to 3, corresponding to more aggressive optimization.
The most aggressive optimizations include noise adaptive qubit mapping and gate cancellation.
This compilation reduces the circuit to one with fewer gates but the same functionality.
Still, such optimization only modestly reduces gate counts in the cases discussed herein.

The IBM Brisbane device is the backend of the transpilation process.
Brisbane supports $127$ qubits and is sufficient for resource estimation.
The prior case of \cref{table:t} transpiled the circuit into a generalized set of gates.
With the Brisbane backend, the transpiler identifies the device-specific gate set, which includes single-qubit X, RZ, and SX gates and the two-qubit Echoed Cross-Resonance (ECR) gate.
ECR gates and a series of single-qubit rotations are applied to implement a CNOT gate.
Thus, reducing the ECR gate count is commensurate with a reduced CNOT count.
The transpilation, optimization, and resource estimation processes are both performed on a local classical device before simulation.

\Cref{table:brisbanecounts} illustrates the resource estimation for a $16 \times 16$ lattice size case.
All algorithms shown in the table, including the QLBM algorithm by \citet{budinski2021quantum}, are optimized via the Qiskit transpiler before resource estimation.
\Cref{table:brisbanecounts} shows that, after optimization, we see a 33\% reduction in ECR gates.
Of the total ECR gates applied, \cref{table:brisbanecounts} shows that the boundary conditions make up 44\% of the ECR gate counts in the QLBM-frugal algorithm.
It remains to be shown whether the reduction in two-qubit gates offsets the cost of computing these bounds on a classical device.
Doing so involves accounting for the lattice size, hardware capabilities, and error rates of the gates.
By computing the results of the stream function and vorticity circuits concurrently, the circuit depth is reduced by 41\%.

\begin{table}
    \centering
    \small
    \begin{tabular}{r r r}
     & ECR Gates & Circuit Depth \\ \midrule
    Single-circuit QLBM & 43\phantom{.0} & 135 \\
    Stream function & 7.6 & 29 \\
    Vorticity & 21\phantom{.0} & 80 \\
    Stream function without boundaries & 7.5 & 28 \\
    Vorticity function without boundaries & 5.0 & 17 \\
    \end{tabular}
    \caption{
        Quantum resource estimation, in units of $10^4$, for a D2Q5 algorithm with lattice size $16 \times 16$ following transpiling with optimization level $3$ on the IBM Brisbane backend.}
    \label{table:brisbanecounts}
\end{table}

\begin{figure}[h!]
    \centering
    \includegraphics{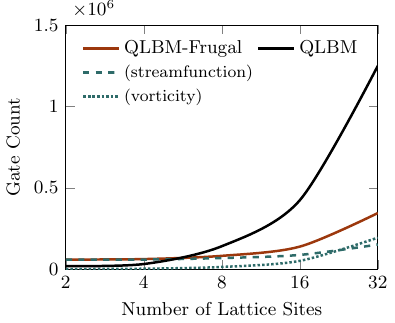}
    \caption{
        Comparison of costly two-qubit gate counts on the D2Q5 algorithm for the proposed two-circuit approach against \citet{budinskiquantumalgorithmadvection2021}'s algorithm.
        Transpiled on the IBM Brisbane backend.
        The horizontal axis ``Number of Lattice Sites'' scales logarithmically.
    }
    \label{f:ecrscaling}
\end{figure}

We now consider how the resource estimation changes with respect to the lattice size.
First, the growth in ECR gate counts is observed in \cref{f:ecrscaling} as the algorithm scales from a $2 \times 2$ to $32 \times 32$ lattice size.
\Cref{f:ecrscaling} shows the number of two-qubit gates required by \citet{budinski2021quantum}'s QLBM algorithm increases more rapidly than that of the proposed QLBM-frugal algorithm.
The number of ECR gates is on the order of one million, implying the difference in cumulative error between the initial and proposed algorithm is notable.
We obtained a runtime estimate for each transpiled circuit corresponding to these lattice sizes via the Qiskit scheduler.
These are calculated based on the execution time of the gates on the specified backend.
These estimates do not account for the encoding and readout costs, which are discussed in \cref{s:limitations}.

\begin{figure}[h!]
    \centering
    \includegraphics{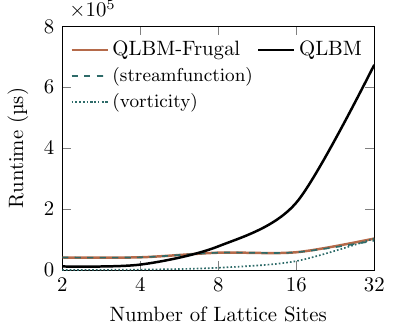}
    \caption{
        Runtime estimation for the proposed D2Q5 QLBM-frugal algorithm without quantum bounds, compared against the QLBM algorithm proposed by~\citet{budinskiquantumalgorithmadvection2021}.
        The circuit is transpiled with optimization level $3$ on the IBM~Brisbane backend.
    }
    \label{f:runtimescaling}
\end{figure}

\Cref{f:runtimescaling} shows the runtime of the respective algorithms.
The stream function and vorticity circuits are presumed to run concurrently, indicating that the QLBM algorithm's runtime scales with the fastest-growing circuit.
Note that the runtime of all of the circuits considered exceeds the coherence time of the IBM~Brisbane hardware, as with most available quantum hardware at present.
These coherence times are on the order of several hundred microseconds.
Thus, results can not be reliably computed on a real device for even the smallest number of lattice sites. 
Furthermore, any classical simulations that seek to mimic the error rates of available hardware will return only noise.
Nonetheless, \cref{f:runtimescaling} demonstrates that running QLBM-frugal in parallel achieves a reduction in runtime when compared to the work by \citet{budinskiquantumalgorithmadvection2021}.

\section{Limitations of current work}\label{s:limitations}

\subsection{Hardware limitations}

Current quantum devices have high noise floors and low qubit counts.
LBM problems are often large, requiring many qubits and so beyond the capabilities of current devices.
The results here follow from quantum simulation.
Concurrently executing the stream function and vorticity circuits doubles the number of required qubits if they are "parallelized" on the same quantum device.
Instead, the quantum work can be distributed across multiple quantum processors, though we limit the present work to single-device analysis.

While the qubit count is indeed doubled, the depth of each circuit is reduced from the initial algorithm proposed by \citet{budinskiquantumalgorithmadvection2021}.
This strategy reduces the overall runtime and two-qubit gate count.
However, the estimated runtime exceeds the coherence time of available hardware by several orders of magnitude.
The presented method still depends on quantum simulators, even for small problems.

\subsection{Encoding and readout costs}

The algorithm of \citet{budinskiquantumalgorithmadvection2021} and the presented two-circuit approach begin with an arbitrary state.
This approach assumes that the state has been encoded in the quantum RAM or an equivalent qubit set.
Unless the amplitudes are roughly uniform, the encoding process is costly.
For some quantum algorithms, this could reduce the expected quantum advantage over their classical counterparts to a polynomial one~\citep{Aaronson_2015}.
In the Qiskit implementation of state preparation, the resource cost scales exponentially with respect to the number of qubits~\citep{Shende_2006}.
Thus, appreciating the impacts of encoding can inform the application of QLBM algorithms.

The final readout process depends on the final solution state.
Quantum state tomography obtains (reads out) the state of the quantum device.
One requires $2^d - 1$ measurements to determine the quantum state of a $d$-dimensional Hilbert space~\citep{Hai_Ho_2023}.
This exponential readout cost can meaningfully impact the runtime and required resources.
This cost motivates an implementation that bypasses exponential readout costs or admits a heuristic-equipped final state for tomography.
More efficient quantum state tomography methods exist in special cases, and the topic remains an active research problem~\citep{Hai_Ho_2023,Quek_Fort_Ng_2021,Rocchetto_2019,Cramer_2010,Aaronson_Learnability}.

\section{Conclusion}\label{s:conclusions}

This work presents improvements on the quantum lattice Boltzmann method for solving the two-dimensional Navier--Stokes equations.
In particular, we propose a QLBM algorithm for concurrently circuit computation of the stream function and vorticity.
These modifications give rise to the possibility of solving the Navier--Stokes equations via parallelization or distributed computing.
We first verified the solution returned by the proposed QLBM-frugal algorithm against classical LBM methods.
We selected the lid-driven cavity problem for algorithm verification, implementing both QLBM and QLBM-frugal using Qiskit.
The error between the results is negligible on ideal quantum devices, indicating the algorithm is practical on fault-tolerant devices.

Moreover, we show QLBM-frugal achieves a reduction of the two-qubit gate count from the previous single-circuit implementation of an otherwise similar algorithm~\citep{budinski2021quantum}. 
Quantum devices are bottlenecked in part by limitations due to error and environmental interference.
Reducing two-qubit gates will improve the algorithm's accuracy.
We demonstrate a 33\% reduction in the two-qubit ECR gate count following optimization and transpilation on the IBM~Brisbane device for the $16 \times 16$ lattice and a 35\% reduction in CNOT gate count when the algorithm is transpiled relative to a general gate set on the larger $64 \times 64$ lattice, corresponding to a reduction in the total circuit depth. 
Concurrent computation of the stream function and vorticity circuits reduces the depth on the $16 \times 16$ lattice size case by 41\%.
This reduction is significant: even for small problems, nearly one million individual gates are required to implement the circuit on quantum hardware.
This reduction removes $\mathcal{O}(10^5)$ qubit rotations and gates from the computation for each iteration and reduces the accumulated error associated with continued gate application.

The reduction in ECR gates and depth extends to cases beyond the $16\times 16$ lattice size.
We show that the QLBM-frugal resource requirements for lattice sizes ranging from $2 \times 2$ to $32 \times 32$ grow at a slower rate than that of the QLBM algorithm.
The expected number of qubits scales logarithmically with the lattice size, $O(\log{M})$, slower than the linear growth demonstrated in \citet{yepezQuantumLatticegasModel2002}.
The runtime of the present work, QLBM-frugal, scales more slowly than the traditional QLBM case. 
While we are far from being capable of implementing such a large circuit on current NISQ hardware, the changes make the circuit more feasible and less error prone on an ideal device relative to its predecessors.

The encoding and readout costs still remain a bottleneck for the algorithm.
The processes must be performed at each iteration to input the prior state of the system and 
extract the information following each computation.
This process is costly, and the processes scale exponentially with respect to the qubit number.
The current work does not address read-in/out costs, though a full QLBM strategy on NISQ devices requires this attention.

\section*{Acknowledgments}

We thank Professors J.~Yepez and L.~Budinski for fruitful discussions of this work.
This work was supported in part by DARPA grant number HR0011-23-3-0006, the Georgia Tech Quantum Alliance, and a Georgia Tech Seed Grant.
This research used resources of the Oak Ridge Leadership Computing Facility, which is a DOE Office of Science User Facility supported under Contract DE-AC05-00OR22725.

\bibliography{ref.bib}

\end{document}